\documentclass[preprint,preprintnumbers,amsmath,amssymb]{revtex4}

\usepackage{graphicx}
\usepackage{dcolumn}
\usepackage{bm}

\begin{document}
\title{Second sound in 2D Bose gas: from the weakly interacting to the strongly interacting regime}
\author{Miki Ota and Sandro Stringari}
\affiliation{INO-CNR BEC Center and Dipartimento di Fisica, Universit\`a di Trento, 38123 Povo, Italy }
\date{\today}
\begin{abstract}
Using Landau's theory of two-fluid hydrodynamics, we investigate first and second sound propagating in a   two-dimensional Bose gas. We study the temperature and interaction  dependence of both sound modes and show that their behaviour exhibits  a deep qualitative change as the gas evolves from the weakly interacting to the strongly interacting regime. Special emphasis is given to the jump of both sounds at the Berezinskii-Kosterlitz-Thouless transition, caused by the discontinuity of the superfluid density. We find that the excitation of second sound through a density perturbation becomes weaker and weaker as the interaction strength increases as a consequence of the decrease of the thermal expansion coefficient. Our results can be relevant for future experiments on the propagation of sound in the BEC side of the BCS-BEC crossover of a 2D superfluid Fermi gas.
\end{abstract}

\maketitle


\par
Superfluidity is one of the most remarkable manifestation of quantum physics at the macroscopic level occurring in diverse systems, from cold atomic gases \cite{Pitaevskii2016, Pethick2001, Bloch2008, Giorgini2008} to neutron stars \cite{Dean2003}. Below the critical temperature $T_c$ at which the phase transition occurs, the system exhibits a two fluid behaviour \cite{Landau1941, Landau1987}, characterized by a mixture of a normal component, behaving as a viscous fluid, and a superfluid component, moving without friction. In these systems, the superfluid density plays a key role for the understanding of related phenomena, such as the frictionless flow of the superfluid \cite{Onofrio2000, Miller2007} and the formation of quantized vortices \cite{Matthews1999, Madison2000, Zwierlein2005}. In a weakly interacting 3D Bose gas the superfluid density is directly related to the experimentally accessible Bose-Einstein condensate fraction. However, this is no longer true for strongly interacting systems, such as $^4 \text{He}$ or for the unitary Fermi gas, where one does not have a straight correspondence between the superfluid and the condensate densities. The situation is even more challenging in 2D systems, where Bose-Einstein condensation is ruled out at finite temperature, as a direct consequence of the Hohenberg-Mermin-Wagner theorem \cite{Hohenberg1967, Mermin1966}. For these systems, a promising way to investigate superfluidity and to identify the value of the superfluid density concerns the measurement of second sound \cite{Pitaevskii2015, Donnelly2009}. This phenomenon arises from the two-fluid nature of the system and corresponds to a wave propagation of the normal  and superfluid components with opposite phase, with a speed of sound  directly related to the superfluid density. Experimentally, the way to probe second sound  depends in a crucial way on the nature of the system. While in $^{4}\text{He}$ or in the unitary Fermi gas second sound is essentially an entropy oscillation, and is conveniently  excited through a thermal perturbation \cite{Taylor2009, Hou2013, Arahata}, the situation drastically changes for a weakly interacting Bose gas where the coupling between entropy and density oscillations becomes important, because of the large value of the thermal expansion coefficient, allowing for  the excitation of second sound through  a density perturbation \cite{Verney2015, Meppelink}. Recently, second sound was observed in the  unitary Fermi gas, yielding first information on the temperature dependence of  the superfluid density \cite{Sidorenkov2013}. First experiment on the propagation of second sound  in a weakly interacting  2D Bose gas has been also recently become available \cite{DalibardPC}.
\par
In this paper, we study the nature and experimental accessibility of first and second sound in 2D Bose gases, exploring the transition from the weakly interacting to the strongly interacting regimes. The former case was  already investigated in \cite{Ozawa2014}, pointing out the occurrence of discontinuities of both sound modes at the Berezinskii-Kosterlitz-Thouless (BKT) transition \cite{Berezinskii1972, Kosterlitz1973}, as a direct consequence of the jump of the superfluid density. In the present work we extend the investigation  to the strongly interacting case, which corresponds experimentally to the BEC regime of a 2D Fermi gas \cite{Levinsen2015, Bertaina2011}. In particular we show that the discontinuity of the first sound  velocity becomes less and less pronounced in the strongly interacting regime, while it remains sizable in the case of second sound. Since in two-dimensional systems the thermodynamic quantities derivable from the equation of state do not show any discontinuity at the phase transition, the experimental measurement of second sound would also provide a unique way to observe directly the BKT phase transition. This is particularly interesting for 2D Fermi gas, where the recent observation of the BKT jump, based on the measurement of pair momentum distribution \cite{Murthy2015}, already simulated a debate in the litterature \cite{Matsumoto2016}. 


\par
Throughout this paper we consider a two dimensional gas, where the third direction is assumed to be blocked. In practice this condition is well satisfied in experiments \cite{Boettcher2016, Fenech2016, Makhalov2014, Desbuquois2012}. We also set $\hbar = k_B =1$ for simplicity. We start our investigation from Landau's two-fluid hydrodynamic equations, describing the finite-temperature dynamics of a uniform system in the superfluid phase. The equations assume local thermodynamic equilibrium, ensured by collisions. In the limit of small amplitude oscillations, the linearized  Landau equations take the form 
\begin{equation}\label{eq.1}
\frac{\partial^2 n}{\partial t^2} = \nabla^2 P,
\end{equation}
\begin{equation}\label{eq.2}
\frac{\partial^2 \bar{s}}{\partial t^2} = \frac{n_s \bar{s}^2}{n_n} \nabla^2 T,
\end{equation}
where $n=n_n+n_s$ is the total atom density, given by the sum of the normal density $n_n$ and the superfluid density $n_s$. $P$ is the pressure, $\bar{s}$ and $T$ are the entropy at constant volume per particle and the temperature, respectively. By looking for plane-wave solutions and using general thermodynamic relations, Eqs. \eqref{eq.1} and \eqref{eq.2} give rise to the the quartic equation,
 \begin{equation}\label{eq.3}
c^4 - \left[ \frac{1}{mn\kappa_s} + \frac{n_s T \bar{s}^2}{mn_n\bar{c}_v} \right] c^2 + \frac{n_s T \bar{s}^2}{mn_n\bar{c}_v} \frac{1}{mn\kappa_T} = 0,
\end{equation}
for the sound velocity, where $m$ is the mass of atom, $\bar{c}_v$ the specific heat at constant volume per particle, $\kappa_s$ and $\kappa_T$ are the adiabatic and thermal compressibilities, respectively. Below the critical temperature Eq.\eqref{eq.3} possesses two positive solutions, corresponding to first and second sound.
\par
In this work, all the thermodynamic quantities are calculated using the universal relations (UR) for the weakly interacting 2D Bose gas derived in \cite{Prokofev2001, Prokofev2002, Yefsah2011, Rancon2012}. The theory provides dimensionless universal functions $f_n(x,g)$ and $f_P(x,g$) depending on the variable $x=\mu/T$, with $\mu$ the chemical potential, and on the dimensionless coupling constant $g$. These functions are related to the density and to the pressure of the gas according to
\begin{equation}
f_n (x,g) = \lambda_T^2 n, \qquad f_P(x,g)= \frac{\lambda_T^2}{T} P,
\end{equation} 
where $\lambda_T=\sqrt{2\pi/mT}$ is the thermal de Broglie wavelength, and are related each other by the thermodynamic relation $f_n=\partial f_P/\partial x$.  Starting from these functions one can then derive expressions for all the quantities appearing in Eq. \eqref{eq.3} \cite{Ozawa2014}, namely
\begin{align}\label{eq:thermo UR}
\begin{split}
\bar{s}=2\frac{f_P}{f_n}-x, \qquad \kappa_T = \frac{1}{nT}\frac{f_n'}{f_n}, \qquad \kappa_s = \frac{1}{nT}\frac{f_n}{2f_P} \\
\bar{c}_v = 2\frac{f_P}{f_n}-\frac{f_n}{f_n'}, \qquad \bar{c}_v = \left( 2\frac{f_P}{f_n}-\frac{f_n}{f_n'} \right) 2 \frac{f_p f_n'}{f_n^2},
\end{split}
\end{align}
where $f_n'=\partial f_n / \partial x$, and $\bar{c}_p$ is the specific heat at constant pressure, per particle. Universal relations further provide another dimensionless function $f_s(x,g) = \lambda_T^2 n_s$, from which one can evaluate the superfluid density. In our work, we use the analytical expression for the dimensionless functions provided in \cite{Prokofev2002}. We note that Ref. \cite{Prokofev2002} also provides Monte Carlo values for $f_n$ and $f_P$  and we have verified that both approaches give  practically the same results. 
\par
\begin{figure}[t]
\begin{center}
\includegraphics[width=0.4\textwidth]{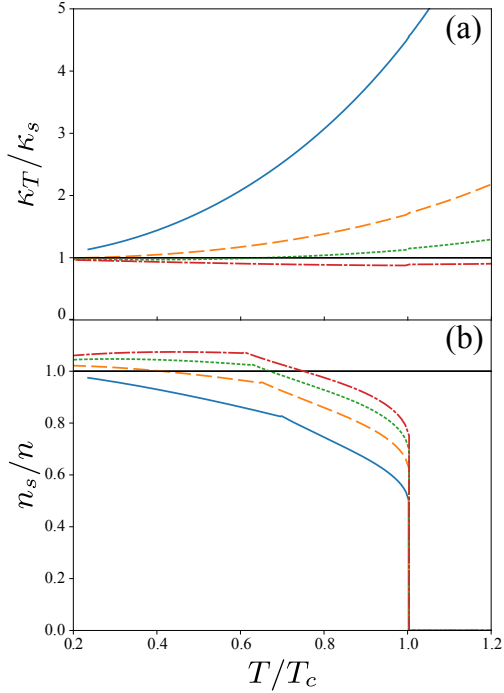}
\caption{ (a) Ratio of isothermal and adiabatic compressibilities $\kappa_T/\kappa_s$ for different values of $g$. From top to bottom, $g=0.1$ (solid line), $g=0.5$ (dashed line), $g=1$ (dotted line), $g=1.5$ (dashed-dotted line).%
(b) Superfluid density fraction $n_s/n$ for different values of $g$. The values of $g$ are the same as in panel (a). The unphysical kinks observed for $T\simeq 0.6T_c$ in the superfluid density is due to the analytical treatment of the dimensionless functions in the universal relations approach \cite{Prokofev2002}. The black solid line is an eye-guide for (a) $\kappa_T/\kappa_s = 1$ and (b) $n_s/n=1$.} 
\label{fig:UR}
\end{center}
\end{figure}
\par
Figure \ref{fig:UR}(a) shows the ratio of thermal and adiabatic compressibilities as a function of temperature, for different values of the coupling constant. The figure shows that  this ratio, which also fixes the value of the thermal expansion coefficient (see Eq. \eqref{eq.7} below), decreases  as the repulsive interaction between bosons becomes stronger. However, from thermodynamic principles, the ratio $\kappa_T/\kappa_s$ can not be smaller than $1$, and Fig. \ref{fig:UR}(a) shows a clear failure of the predictions based on the UR for $g \geq 1$.  In Fig. \ref{fig:UR}(b) we show the superfluid density fraction $n_s/n$ for the same values of the coupling constant. Again we see another failure of the universal relation which predicts a value for the ratio $n_s/n$ larger than 1 at low temperature if the coupling constant $g$ is large enough. This failure is the consequence of the fact that the UR correctly describe only the  fluctuating region  near the critical point\cite{Prokofev2001, Prokofev2002}. As the interaction increases this  region around $T_c$ shrinks, reducing the region of applicability of the  UR approach, although it allows for a good estimate of  $T_c$ also for large values of $g$, as confirmed by the comparison with  ab initio Quantum Monte Carlo calculations  \cite{Pilati2008}. For the above reasons in  the following we will limit our theoretical analysis, based on the predictions of the UR approach  to values $g \leq 1$. We briefly note that, $g \simeq 0.1$ is a typical value of coupling constant for a dilute 2D Bose gas \cite{Desbuquois2012}, and values $g \lesssim 2$ correspond to the BEC regime of a 2D Fermi gas \cite{Boettcher2016, Makhalov2014}, where the system is expected to behave physically like a gas of bosonic dimers \cite{Note1}.
\par
\begin{figure}[t]
\begin{center}
\includegraphics[width=0.8\textwidth]{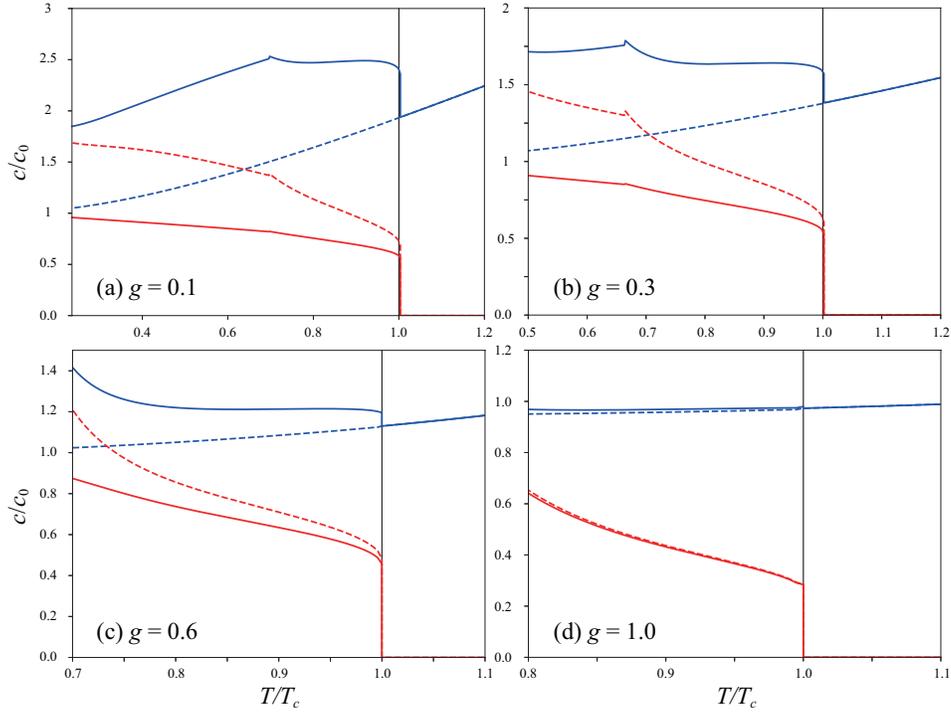}
\caption{First and secound sound as a function of temperature for different values of $g$. The blue and red solid lines correspond to first and second sound calculated from Eq. \eqref{eq.3}, respectively. The blue and red dashed lines are the approximated form of the first and second sound for small thermal expansion coefficient, given by Eq. \eqref{eq.8}.} 
\label{fig:sound0} 
\end{center}
\end{figure}
\par
Figure \ref{fig:sound0} shows the first and second sound obtained by solving Eq. \eqref{eq.3} (solid line), for different values of $g$ (the results for $g=0.1$ were already reported in \cite{Ozawa2014}). The velocities are calculated for a fixed value of the total density  and are expressed in units of the zero temperature Bogoliubov sound velocity $c_0 = \sqrt{gn}/m$. As one can see, both sound velocities show a jump at the transition temperature. This behaviour, originating from the BKT universal jump of the superfluid density, will be studied in details in the following. In order to understand the evolution of the sound modes with the coupling constant one notices that  if the thermal expansion coefficient $\alpha = -\frac{1}{n} \frac{\partial n}{\partial T} \vert_P$ satisfies the condition
\begin{equation}\label{eq.7}
\alpha T = \left( \frac{\kappa_T}{\kappa_s} -1 \right) \ll 1,
\end{equation}
the two solutions of Eq. \eqref{eq.1} and \eqref{eq.2}  take the form of wave equations for the density and for the entropy respectively, the corresponding sound velocities being given by
\begin{equation}\label{eq.8}
c_{10}^2 = \frac{1}{mn\kappa_s}, \qquad c_{20}^2=\frac{n_s T \bar{s}^2}{mn_n\bar{c}_p}.
\end{equation}
\begin{figure}[t]
\begin{center}
\includegraphics[width=0.6\textwidth]{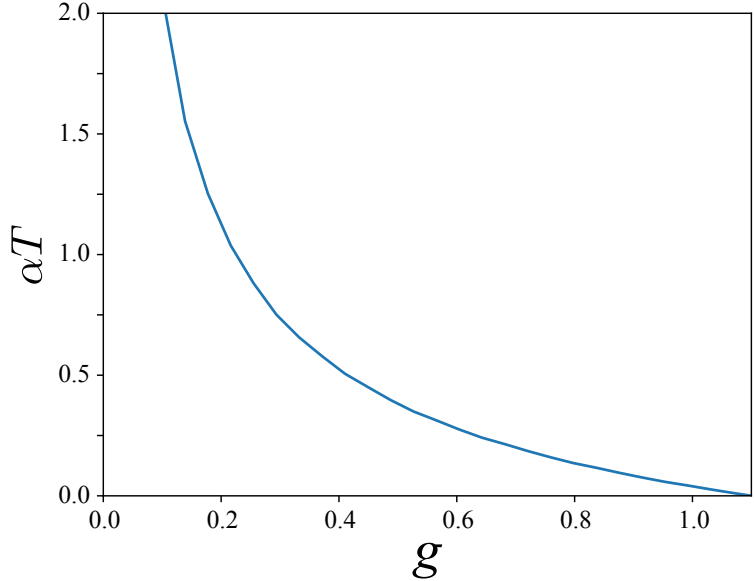}
\caption{Thermal expansion coefficient $\alpha T$ at $T=0.8T_c$ as a function of the 2D coupling constant $g$.} 
\label{fig:alphaT} 
\end{center}
\end{figure}
The figure shows  that the calculated velocities strongly deviate from Eq. \eqref{eq.8} (shown as dashed-line in Fig. \ref{fig:sound0}) for panels (a) and (b), revealing the strong coupling between the density and entropy modes in the highly compressible regime where  the condition  $\alpha T \ll 1$ is violated. Figure \ref{fig:alphaT} shows that as the coupling constant increases, the gas evolves from a weakly interacting   to a strongly interacting behaviour, becoming less compressible. As a consequence  Eq. \eqref{eq.8} becomes more and more accurate, as shown on panels (c) and (d). The transition between the weakly interacting and the strongly interacting regime is then expected to take place for values of the 2D coupling constant  corresponding to  $g\sim 0.5$. This regime is not too far from experimental achievability in the BEC side of the BEC-BCS crossover in 2D superfluid Fermi gases. It is worth noticing that the already mentioned unphysical violation of the thermodynamic  relation  $\kappa_T/\kappa_s \ge 1$ predicted by the use of universal relations for large values of the coupling constant, has little effect on the sound speeds, while the violation of the condition $n_s \le n$  has dramatic unphysical  consequences due to the resulting negativity of the normal density. The proper estimate of the sound velocities in the strongly interacting regime should then be based on more realistic estimates of  the superfluid density. Accurate calculations of the superfluid density as well as of the relevant thermodynamic functions of 2D Fermi gases, based on quantum Monte Carlo simulations \cite{Pilati2008, Anderson2015} or many-body theories \cite{He2015, Mulkerin2017, Bighin2016},  would in particular allow for a safer evaluation of the sound velocities along the whole BCS-BEC crossover. 
\par
\begin{figure}[t]
\begin{center}
\includegraphics[width=0.8\textwidth]{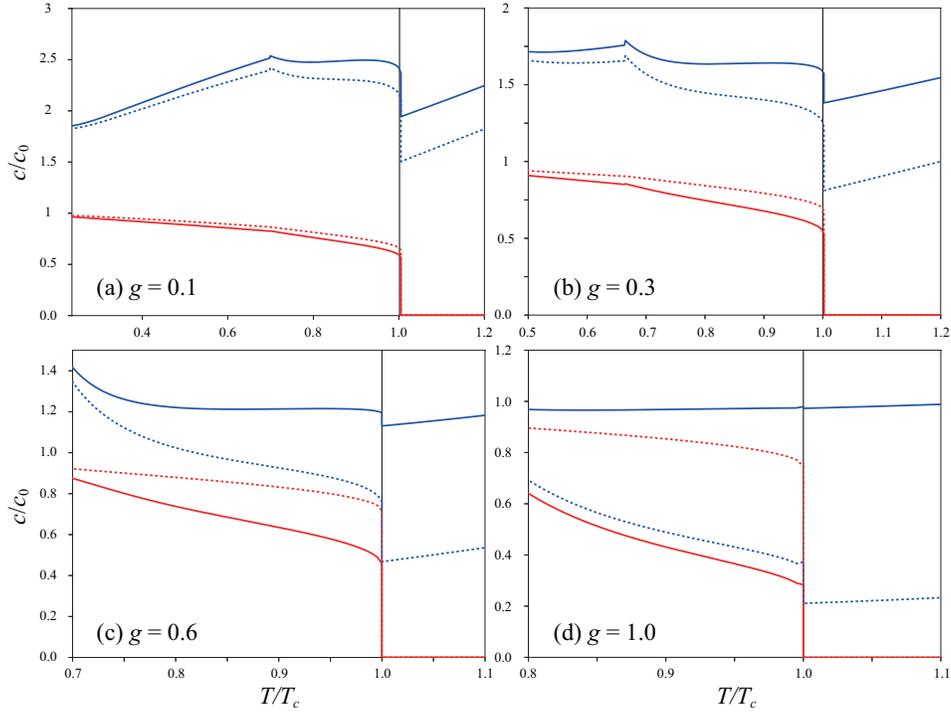}
\caption{First and secound sound as a function of temperature for different values of $g$. Solid lines are the same as Fig. \ref{fig:sound0}. The blue and red dotted lines are the approximated form of the first and second sound for weakly interacting Bose gas, given by Eq. \eqref{eq.9}.} 
\label{fig:soundWI} 
\end{center}
\end{figure}
\par
In the case of very dilute Bose gases, an accurate approximated solution of Eq. \eqref{eq.3} is obtained  by replacing all the relevant  thermodynamic quantities, except the isothermal compressibility and the superfluid density, with the values predicted by the ideal Bose gas \cite{Pitaevskii2015}. Then Eq. \eqref{eq.3} gives,
\begin{equation}\label{eq.9}
c_{1,WI}^2 = \frac{nT\bar{s}^2}{n_n m \bar{c}_v}, \qquad c_{2,WI}^2=\frac{n_s}{n}\frac{1}{mn\kappa_T}.
\end{equation}
Figure \ref{fig:soundWI} shows again the sound velocities for the same values of the coupling constant, but compared this time to Eq. \eqref{eq.9} (dotted-lines). As expected, the approximation successfully describes the exact sound speeds for small $g$ and  becomes less and less accurate as one increases the value of $g$.
\par
While in the 3D case the sound velocities near $T_c$ can be estimated by putting $n_s \rightarrow 0$ leading to Eq. \eqref{eq.8}, this assumption can not be used in  2D because of the presence of the gap. One can however derive a first-order correction to the values of  $c_{10}$ and $c_{20}$ resulting from the solution of   Eq. \eqref{eq.3}, by assuming $\alpha T c_{20}^2/c_{10}^2 \ll 1$. One finds:
\begin{equation}\label{eq.10}
c_{1,BKT}^2 = c_{10}^2 \left( 1 + \alpha T \frac{c_{20}^2}{c_{10}^2} \right), \qquad c_{2,BKT}^2 = c_{20}^2 \left( 1 - \alpha T \frac{c_{20}^2}{c_{10}^2} \right) ,
\end{equation}

\begin{figure}[t]
\begin{center}
\includegraphics[width=0.6\textwidth]{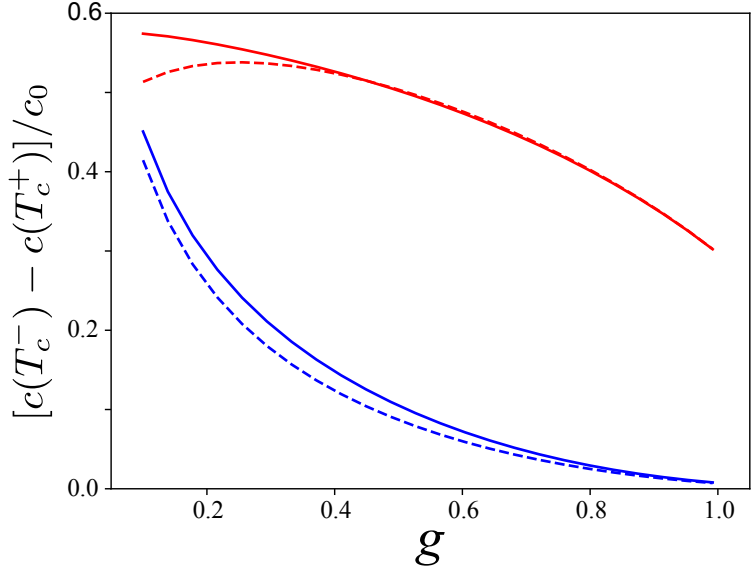}
\caption{BKT jump in sound velocities $c_{BKT}^- - c_{BKT}^+$ as a function of $g$. Velocities jump for first sound (lower solid line) and second sound (blue solid line) obtained from Eq. \eqref{eq.3} (red solid lines) are compared to the approximated expression Eq. \eqref{eq.10} (blue and red dashed lines for first and second sounds, respectively).} 
\label{fig:jump} 
\end{center}
\end{figure}

Results \eqref{eq.10} are expected to be valid near $T_c$, and in particular they correctly describe the  jump $c(T_c^-) - c(T_c^+)$ of the first and second sound velocities when one crosses the critical temperature for  a wide range of values of the coupling constant, as explicitly shown in Fig. \ref{fig:jump}. According to Eq. \eqref{eq.10}, the deviation of the sound velocities from $c_{10}$ and $c_{20}$ near $T_c$ is characterized by the factor $\alpha T c_{20}^2/c_{10}^2$ revealing the crucial role played by the difference between the thermal and the adiabatic compressibilities. This is explicitly shown in Fig. \ref{fig:sound0}(d), where, for large values of $g$, the jump  of  first sound disappears due to the vanishingly small value of thermal expansion coefficient $\alpha$.
\par
\par
\begin{figure}
\begin{center}
\includegraphics[width=0.6\textwidth]{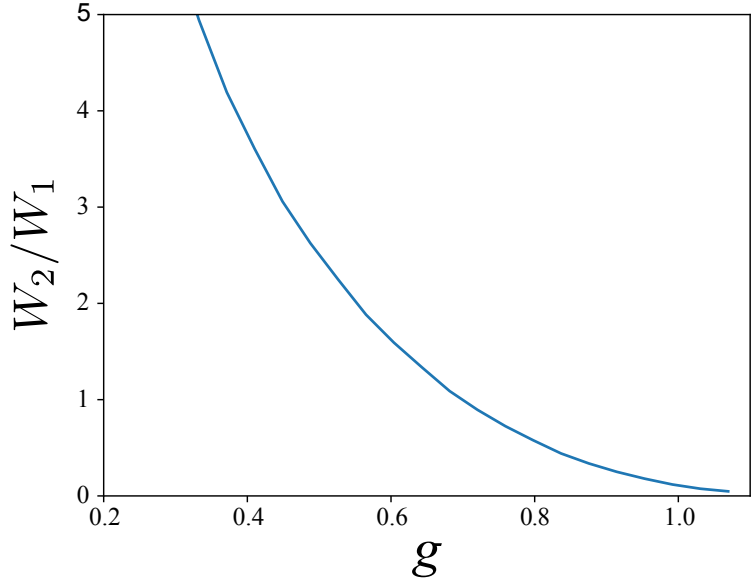}
\caption{Ratio of compressibility sum rule contribution $W_2/W_1$ at $T=0.8T_c$. $W_1$ ($W_2$) is the relative contribution of first (second) sound mode to the compressibility sum rule Eq. \eqref{eq.12}.} 
\label{fig:W2W1}
\end{center}
\end{figure}

As briefly mentionned in the introductory part, it is of highly interest to understand if second sound can be excited using a density probe. Experimentally, this can be achieved using a sudden laser perturbation applied to the center of the trap, or through a sudden modification of the confining potential in the case of a box potential. By assuming that the perturbation acts on  macroscopic length scales, in the linear approximation the induced density fluctuations  are determined by the static polarizability, fixed by the compressibility sum rule \cite{Pitaevskii2016}
\begin{equation}\label{eq.11}
\lim_{\mathbf{q} \rightarrow 0} \int_{-\infty}^\infty d\omega \frac{1}{\omega} S(\mathbf{q},\omega) = \frac{1}{2} n\kappa_T.
\end{equation}
where $S(\mathbf{q},\omega)$ is the dynamical structure factor with wave vector $\mathbf{q}$ and frequency $\omega$. On the other hand, the energy weighted momentum also satisfies the energy-weighted sum rule  $\int_{-\infty}^{\infty} d\omega \omega S(\mathbf{q},\omega) = q^2/(2m)$. Since in the macroscopic limit of small $\mathbf{q}$ one expects that the two sum rules are exhausted by the two sound modes, one can express the relative contribution of each sound mode to the compressibility sum rule Eq. \eqref{eq.11} in the form \cite{Hu2010}
\begin{equation}\label{eq.12}
W_1 = \frac{1-mn\kappa_Tc_2^2}{2m(c_1^2-c_2^2)}, \qquad W_2 = \frac{mn\kappa_Tc_1^2-1}{2m(c_1^2-c_2^2)},
\end{equation}
where we have naturally chosen $c_1 > c_2$. If the ratio $W_2/W_1$ is not too small, second sound can be excited through a density perturbation. We also note that, under the assumption $c_1 \geq c_{10}$, the  thermal expansion coefficient sets the lower bound 
\begin{equation}\label{eq.13}
\frac{W_2}{W_1} \geq \alpha T.
\end{equation} 
Figure \ref{fig:W2W1} shows the ratio of the relative contribution of second and first sound to the compressibility sum rule Eq. \eqref{eq.12}, calculated by solving the Landau equation \eqref{eq.3}.  From the comparison with Fig. \ref{fig:alphaT} we can see that, as expected from Eq. \eqref{eq.13}, the ratio $W_2/W_1$ follows the same evolution as $\alpha T$. This observation explicitly reveals that  the excitation of second sound via a density probe becomes more and more difficult as one increases the value of the coupling constant.
\par
Since the BEC regime of a 2D Fermi gas can be described in terms of  an interacting molecular Bose gas, our results  provide valuable information for the description of this system in a useful range of hopefully experimentally accessible parameters. From this point of view, the most interesting region to explore experimentally  would be around $g \simeq 0.5$, where the ratio $W_2/W_1 \simeq 2$ is still large to allow for the excitation of second sound via a density probe. Such experiments would provide unique information on the value of the superfluid density and on the applicability of the universal relations for 2D Bose gases beyond the weakly interacting regime.
\par
In conclusion we have provided a systematic investigation of the   behavior of second sound in a 2D interacting Bose gas, exploring the transition between the weakly interacting limit to the regime characterized by larger values of the 2D coupling constant $g$.  Second sound is sensitive to the behavior of the superflud density and its measurement can then provide unique information on the effects of superfluidity, a phenomenon of high interest, especially in two dimensions, where the system is characterized by  the Berezinskii-Kosterlitz-Thouless transition. We have shown that the nature of second sound exhibits a deep change as a function of the coupling constant. For small values of $g$ second sound can be identified  as a density wave, of  easy experimental detection. For larger values of $g$, second sound looses its density character and takes the form of a temperature, or entropy wave, in analogy with the behavior exhibited by superfluid helium and by the 3D Fermi gas at unitarity.
\par
A challenging open question is to understand whether  the collisional regime, required to apply the Landau two-fluid hydrodynamic approach, is guaranteed in the  experimentally available conditions. A recent experiment \cite{DalibardPC} on the propagation of sound in a weakly interacting Bose gas confined in a 2D box potential has shown that, differently from the predictions of two fluid hydrodynamic equations, a density wave can propagate at low velocity even above the critical temperature, thereby suggesting that the collisional regime is not guaranteed in this experiment. Due to the finite size $L$ of the box,  the frequency of the lowest mode, of  order  $v/L$, where $v$ is the velocity of sound, may not in fact be enough small compared to the collisional frequency, thereby violating the hydrodynamic condition. More theoretical  work is then needed to better understand whether  sound  can propagate in a 2D Bose gas in the absence of collisions.  

\section*{Acknowledgements}

We thank Giacomo Bighin and Luca Salasnich for stimulating discussions. We also thank Stefano Giorgini and Tomoki Ozawa for useful comments. This work has been supported by the QUIC grant of the European Horizon2020 FET program and by Provincia Autonoma di Trento.



\par

\end{document}